\documentclass[prd,aps,12pt,preprintnumbers]{revtex4}

\usepackage{bm}
\usepackage{amsmath}
\usepackage{yfonts}
\usepackage{epsfig}

\usepackage{graphicx}
\usepackage{rotating}
\usepackage{latexsym}

\newcommand{\bqq}{\begin{equation}}
\newcommand{\eqq}{\end{equation}}
\newcommand{\beq}{\begin{eqnarray}}
\newcommand{\eeq}{\end{eqnarray}}

\newcommand{\nn}{\nonumber\\}
\newcommand{\dif}{{\rm d}}

\newcommand{\EA}{\mcal{E}_{\rm A}}
\newcommand{\EB}{\mcal{E}_{\rm B}}
\newcommand{\Eone}{\mcal{E}_{\rm rad}^{(1)}}
\newcommand{\Etwo}{\mcal{E}_{\rm rad}^{(2)}}

\newcommand{\mW}{\mcal{W}}

\newcommand{\del}{\partial}

\newcommand{\mcal}{\mathcal}

\newcommand{\br}{\bm{r}}

\newcommand{\brq}{\bm{r}_q}
\newcommand{\bvq}{\bm{\upsilon}_q}
\newcommand{\baq}{\bm{a}}

\newcommand{\gq}{\gamma}

\begin{document}

\preprint{BI-TP 2011/22}

\title{On radiation by a heavy quark in ${\cal{N}} =$~4~ SYM}

\author{Rudolf Baier\footnote{ E-mail: {baier@physik.uni-bielefeld.de}}}

\affiliation{
    Physics Department, University of Bielefeld, D-33501 Bielefeld,
Germany}

\vspace{2.5cm}

\begin{abstract}
A short note on radiation by a moving classical particle in ${\cal{N}}~=$~4
 supersymmetric Yang-Mills theory.
\end{abstract}

\maketitle

\section{Introduction}

{In the papers \cite{Athanasiou:2010pv,Hatta:2011gh,Maeda:2007be}
radiation by a pointlike quark in ${\cal{N}} =$~4 supersymmetric Yang-Mills
theory at strong coupling is investigated using the AdS/CFT correspondence in
the supergravity 
approximation \cite{Maldacena:1997re,Gubser:1998bc,Witten:1998qj}.
In this note modifications of the  published radiation pattern  
are suggested, which are consistent with the results in \cite{Mikhailov:2003er}.
This analysis is  motivated by the description of  electrodynamic radiation in
classical electrodynamics \cite{Rohrlich,Schwinger}}.

The important result in the context of radiation by an accelerated charge $e$
is given by the Abraham-Lorentz four-vector force 
\cite{Rohrlich,Dirac,Rohrlich2} in classical 
relativistic electrodynamics \cite{Thirring},
\bqq\label{fED}
f^{\mu} = \frac{2 e^2}{3} ( a^{\nu} a_{\nu} v^{\mu} +  {\dot a}^{\mu}   )~,
\eqq
where the particle velocity is ${\dot x}^{\mu} = 
 v^{\mu} \equiv \frac{d x^{\mu}}{d\tau}$
and the acceleration ${\dot v}^{\mu} = a^{\mu} \equiv \frac{d v^{\mu}}{d\tau}$
with the proper time $\tau$
 for the particle \cite{Eriksen} [The signature used here is $(+---)$].
It is important to note the orthogonality of the force to the velocity, 
\bqq\label{constr}
 v_\mu f^\mu = 0~.
\eqq

\noindent
This force vanishes for uniformly accelerated motion, $f^{\mu} = 0$
 \cite{Rohrlich}.

\section{Classical radiation of accelerated electrons}

In order to set the framework, it is helpful to discuss  radiation in
classical electrodynamics. A usefull approach is found in the 1949
 paper by Schwinger
\cite{Schwinger}.

Assume sources, restricted to a finite domain, which emit radiation. The
four-momentum of the classical electromagnetic field is given in terms of the
energy-momentum tensor by integrating on a hyper-surface

\bqq
P^{\nu} = \oint ~T^{\mu \nu} d\sigma_{\mu} ~.
\eqq

\noindent
Gau{\ss} and Maxwell allow us to rewrite it as

\bqq
P^{\nu} = \int  \partial_{\mu}T^{\mu \nu} d^4 x
= \int  j_{\mu} F^{\mu \nu} d^4 x~,
\label{eq:P}
\eqq
in terms of the current $j_{\mu} = (\rho, \vec{j})$ and 
the field-strength $F^{\mu \nu}$.

In the following it is important in order to determine  the radiation force
that
the radiation field tensor and the vector potential are introduced
in terms of the retarded and advanced fields \cite{Dirac,Rohrlich2}:

\bqq\label{rad1}
F^{\mu \nu}_{rad} = \frac{1}{2} (F^{\mu \nu}_{ret} -  F^{\mu \nu}_{adv})~.
\eqq  
Replacing in (\ref{eq:P}) $F^{\mu \nu}$ by $F^{\mu \nu}_{rad}$
gives $P^{\nu}_{rad}$, which for point-like charges satisfies (compare to 
(\ref{constr}))
\bqq
v_\nu \frac{d P^{\nu}_{rad}}{d \tau} \propto 
 v_{\mu} F^{\mu \nu}_{rad} v_\nu = 0~.
\eqq

\noindent
Using current conservation $\partial_{\mu} j^{\mu} = 0$ and introducing the
vector potential
\bqq\label{rad2}
 A^{\mu}_{rad} = \frac{1}{2}~(A^{\mu}_{ret} - A^{\mu}_{adv})
\equiv (\phi, \vec{A})~,
\eqq
one obtains the power

\bqq\label{power}
\frac{d P^0_{rad}}{dt} =
\int~[\vec{j} \cdot \frac{\partial \vec{A}}{\partial t}  - 
 \rho ~\frac{\partial \phi}{\partial t}]~ d^3 x ~
+ \frac{d}{dt}\int \rho ~\phi ~d^3 x~.
\eqq

\noindent
In \cite{Schwinger}, eq.~(I.17), Schwinger discards the second term
 of this formula,
which has the form of a total time derivative. 
 
\noindent
The radiation vector potential \cite{Schwinger} is expressed by

\bqq
A^{\mu}_{rad} (t, \bm{x}) = \frac{i}{2 \pi}~\int
\exp{[i \omega (\bm{n} \cdot (\bm{x} - \bm{x'}) - (t -t'))]}
~j^{\mu}(t', \bm{x'}) ~d^3 x' dt' ~\omega d\omega \frac{d \Omega}{4 \pi}~.
\eqq

\noindent
A point-particle current is assumed
\bqq
j^{\mu} = e (1, \bm{v}(t) = \frac{d \bm{x}}{dt}~)~ \delta (\bm{x} -
\bm{R}(t))~.
\eqq

\noindent
The integrals in (\ref{power}) are as follows:

\bqq
\int  \rho ~ \phi~d^3x = e^2 \int~ \delta' (t' - t +
\bm{n}\cdot(\bm{R}(t) - \bm{R}(t')) ~dt' d\Omega =
~ - e^2 \int \frac{d \Omega}{4 \pi} ~\frac{\bm{n}\cdot \bm{a}}
{\xi^3}~,
\eqq 
with $\xi = {(1 - \bm{n} \cdot \bm{v}~)} $, $\bm{a} = \frac{d\bm{v}}{dt}$,
and from eq. (I.41) of \cite{Schwinger}
 
\beq
&\int&~ [\vec{j} \cdot \frac{\partial \vec{A}}{\partial t}  - 
 \rho ~\frac{\partial \phi}{\partial t}]~ d^3 x ~ \nn
&=& e^2 \int \frac{d \Omega}{4 \pi}~ 
[ \frac{\bm{a}^2}{\xi^3} + 2 \frac{\bm{n} \cdot \bm{a}~ \bm{v} \cdot
  \bm{a}}{\xi^4}
- \frac{(\bm{n} \cdot \bm{a})^2}{\gamma^2 \xi^5}] \nn
&+& \frac{d}{dt} e^2 \int \frac{d \Omega}{4 \pi} ~
[ - \frac{\bm{v} \cdot \bm{a}}{\xi^3} 
+ \frac{\bm{n} \cdot \bm{a}}{\gamma^2 \xi^4}]~,
\label{schw}
\eeq
with $\gamma = \frac{1}{\sqrt{1 - \bm{v}^2}}$,
 and $v^\mu = (\gamma, \gamma \bm{v})$.

\noindent
Finally, using the notion of emitted power and a Schott type term
(for example, in the notation of \cite{Galtsov:2004qz}), 
the result of the angular radiation pattern is

\bqq 
\frac{d P^0_{rad}}{dt~ d\Omega} \equiv P_{rad}(\bm{n},t)
= P_{emitt}(\bm{n},t) + P_{Schott}(\bm{n},t)~,
\eqq
where
\bqq\label{emitt}
 P_{emitt}(\bm{n},t) =  \frac{e^2}{4 \pi}~ 
[ \frac{\bm{a}^2}{\xi^3} + 2 \frac{\bm{n} \cdot \bm{a}~ \bm{v} \cdot
  \bm{a}}{\xi^4}
- \frac{(\bm{n} \cdot \bm{a})^2}{\gamma^2 \xi^5}]~,
\eqq 
and
\bqq\label{Schott}
 P_{Schott}(\bm{n},t)~ = \frac{e^2}{4 \pi}~ \frac{d}{d t}
[ - \frac{\bm{v} \cdot \bm{a} + \bm{n} \cdot \bm{a}}{\xi^3} 
+ \frac{\bm{n} \cdot \bm{a}}{\gamma^2 \xi^4}]~.
\eqq

\noindent
Indeed there are two terms contributing to the radiation power.
Schwinger \cite{Schwinger}  claims that only the first one 
$ P_{emitt}(\bm{n},t)$, the one denoting the emission
should be retained. It has the characteristics of an irreversible energy transfer. 
The second one in the form of a total time derivative is reversible in nature.

Following Jackson \cite{Jackson} to obtain the radiated energy density of a charged
particle one starts from
the  large distance - $1/R$ contribution of the Li$\acute e$nard-Wiechert
electric field
\beq\label{Efield}
 \bm{E}_{\rm rad} &=&
\frac{e}{ R}\, 
\frac{\bm{n} \wedge [(\bm{n} - \bm{v}) \wedge
 \bm{a}]}{(1 - \bm{n} \cdot \bm{v})^3} \nn
&=&
 \frac{e}{ R}\,
 \left[- \frac{\baq}
 {(1 - \bm{n} \cdot \bm{v})^2}
 +\frac{(\bm{n} \cdot \baq)(\bm{n} - \bm{v})}
 {(1 - \bm{n} \cdot \bm{v})^3}\right],
\eeq
to obtain from

\bqq
 \mcal{E}_{\rm vector} = \frac{1}{8 \pi}\left(\bm{E}^2 + \bm{B}^2\right)~,
 \eqq 
with
$|\bm{B}_{\rm rad}| = |\bm{E}_{\rm rad}|$

 \bqq\label{Evec}
 \mcal{E}_{\rm vector} =
 \frac{e^2}{4 \pi R^2}
 \left[\frac{\baq^2}{(1 - \bm{n} \cdot \bm{v})^4} + 2\,\frac{(\bm{v} \cdot \baq)
 (\bm{n}\cdot\baq)}{(1 - \bm{n} \cdot \bm{v})^5} -\frac{(\bm{n}\cdot\baq)^2}
 {\gq^2(1 - \bm{n} \cdot \bm{v})^6} \right].
 \eqq

\noindent
There is agreement between 
\bqq
R^2 (1 - \bm{n} \cdot \bm{v})~ \mcal{E}_{\rm vector} 
= P_{emitt}(\bm{n},t) ~.
\eqq

\noindent
Integrating the angular dependence 
(see e.g. the useful integrals in  Appendix A in \cite{Hatta:2011gh})
one obtains

\bqq\label{eq:twenty} 
\gamma \frac{d P^0_{rad}}{d t} = \frac{d P^0_{rad}}{d \tau} =
\frac{2 e^2}{3}~ \gamma^4 [\bm{a}^2  + \gamma^2 (\bm{v} \cdot \bm{a})^2]
\gamma
- \frac{2 e^2}{3} \frac{d}{d \tau}  [\gamma^4 (\bm{v} \cdot \bm{a})]~,
\eqq 
which,  with $a^{\mu} = (\gamma^4 \bm{v} \cdot \bm{a}, \gamma^2 \bm{a} + \gamma^4
(\bm{v} \cdot \bm{a}) \bm{v})$,
 can be written as
\bqq
\frac{d P^0_{rad}}{d \tau} = - f^0 = -
\frac{2 e^2}{3} ~[~ a^\mu a_\mu v^0 + {\dot{a}}^0~]~,
\eqq
i.e. the zero component of the (negative) relativistic Abraham-Lorentz vector 
force (\ref{fED})
\cite{Rohrlich,Dirac,Thirring,Galtsov:2004qz},
\beq
f^\mu &=& f^\mu_{emitt} + f^\mu_{Schott}  
\nn
 &=& \frac{2 e^2}{3} [ (\frac{d^2 x}{d \tau^2})^2~ \frac{{d x}^\mu}{d \tau}
 + \frac{{d^3 x}^\mu}{d \tau^3}]~.
\eeq
The first term represents an irretrievable loss of energy, the second, the
Schott contribution, is a
total time differential, which contributes nothing to an integral by $d \tau$,
 when the initial value of $a^\mu$ is returned to at the end \cite{Thirring}.

In summary, as Schwinger stated already in 1949, only the 
spectrum for the irreversible transfer
$P_{emitt}(\bm{n},t)$ is the relevant one for discussing the emitted 
radiation, and
therefore should be retained. It is consistent with the derivation via the
radiative electric field (\ref{Efield}) as given in \cite{Jackson}.

\section{Classical ${\cal{N}} =$~4  SYM radiation}

In order to calculate the radiation power one has to add to the vector part
a contribution due to a massless  scalar field $\chi$
 \cite{Athanasiou:2010pv,Hatta:2011gh},
and the replacement $e^2 \rightarrow e_{eff}^2 = \frac{\lambda}{8 \pi}$.
This leads to 
\bqq
\partial_{\mu}T^{\mu \nu}_{scalar} = j_{\chi} \partial^{\nu} \chi~,
\eqq 
with the current
\bqq
j_{\chi} = \rho_{\chi} = e_{eff}~\sqrt{1- v^2} ~\delta(\bm{x} - \bm{R}(t))~.
\eqq
This scalar contribution leads to
\beq
P_{scalar}(\bm{n},t) &=& \frac{e_{eff}^2}{4 \pi}
[\frac{ \gamma^2 (\bm{v} \cdot \bm{a})^2}{\xi^3}
- 2 \frac{(\bm{v} \cdot \bm{a})(\bm{n} \cdot \bm{a})}{\xi^4}
+ \frac{(\bm{n} \cdot \bm{a})^2}{\gamma^2 \xi^5}] 
\nn
&+&  \frac{e_{eff}^2}{4 \pi} \frac{d}{d t} [\frac{\bm{v} \cdot \bm{a}}{\xi^3}
- \frac{\bm{n} \cdot \bm{a}}{\gamma^2 \xi^4}]~,
\eeq
(see also \cite{Athanasiou:2010pv,Hatta:2011gh}).
Adding the vector parts (\ref{emitt}) and (\ref{Schott}) from the previous section
the weak coupling angular spectrum is given by
\bqq\label{weakP} 
P_{rad}(\bm{n},t) =
\frac{\lambda}{32 \pi^2} ~\frac{\bm{a}^2 + \gamma^2 (\bm{v} \cdot
  \bm{a})^2}{\xi^3}
- \frac{\lambda}{32 \pi^2} \frac{d}{d t} [\frac{\bm{n} \cdot \bm{a}}{\xi^3}]~.
\eqq  
The term for $P_{emitt}(\bm{n},t)$ may also be expressed as
\bqq
 P_{emitt}(\bm{n},t) = \frac{\lambda}{32 \pi^2} 
\frac{\gamma^2 [\bm{a}^2 - (\bm{v} \wedge \bm{a})^2]}{(1- \bm{n} \cdot \bm{v})^3}~.
\eqq

\noindent
Performing the angular integration gives 
\bqq
\int  \gamma P_{rad}(\bm{n},t)~ d\Omega
= \frac{\lambda}{ 8 \pi} \gamma^4 [\bm{a}^2 + \gamma^2 (\bm{v} \cdot
  \bm{a})] \gamma - 
 \frac{\lambda}{ 8 \pi} \frac{d}{d\tau}[\gamma^4 (\bm{v} \cdot \bm{a})]~.
\eqq
Up to the coupling this expression agrees with the one from classical
electrodynamics, given by (\ref{eq:twenty}).
The ${\cal{N}} =$~4 SYM Abraham-Lorentz force
  \cite{Chernicoff:2009re,Chernicoff:2009ff}
in the weak coupling limit reads 
\bqq\label{fpert}
f^{\mu}_{SYM, weak} = \frac{\lambda}{8 \pi}[ a^\nu a_\nu v^\mu + {\dot{a}}^\mu]~,
\eqq
allowing the same interpretation as in classical electrodynamics given above.
$f^{\mu}_{SYM,weak}$ satisfies the constraint (\ref{constr}).

\section{Radiation in  ${\cal{N}} =$~4 SYM at strong coupling}

Based on the work by \cite{Athanasiou:2010pv}~
Hatta et al. \cite{Hatta:2011gh} performed a detailed and transparent 
calculation of the
radiation pattern by a heavy quark in ${\cal{N}} =$~4 SYM at strong coupling,
to be  followed rather closely.
The result consists of two parts for the energy density, to be identified as
\bqq{\label{Pemitt}}
P_{emitt}(\bm{n},t) = \frac{\sqrt{\lambda}}{8 \pi^2}
\frac{\gamma^2 [\bm{a}^2 - (\bm{v} \land \bm{a})^2]}{(1 - \bm{n} \cdot
  \bm{v})^3}~,
\eqq 
and a term in form of a total time derivative
\bqq{\label{Ptt}}
P_{tt}(\bm{n},t) = \frac{\sqrt{\lambda}}{24 \pi^2}
\frac{d}{d t}[\frac{\bm{v} \cdot \bm{a}}{\xi^3} -
\frac{\bm{n} \cdot \bm{a}}{\gamma^2 \xi^4}]~,
\eqq
which is - up to the couplings - the same  given by (\ref{schw})
in classical electrodynamics \cite{Schwinger}.

\noindent
In the notation of \cite{Hatta:2011gh} $P_{emitt}(\bm{n},t) = R^2 \xi 
\Eone(t,{\bf r})$ and $P_{tt}(\bm{n},t) = 
R^2 \xi  \Etwo(t,{\bf r})$.
It is noted in \cite{Hatta:2011gh} 
that integrating the sum of these two terms with respect to $d\Omega$ does not give a proper 
Abraham-Lorentz force \cite{Chernicoff:2009re,Chernicoff:2009ff},
and the constraint (\ref{constr}) is not satisfied,
as it is the case in the weak coupling limit, when compared with (\ref{fpert}).

A possible source of this deficiency may be found that only retarded
contributions for the  radiation are taken into account, instead of following
the prescription given e.g. by (\ref{rad1}) and (\ref{rad2}) in the previous sections.

\noindent
There is no need to repeat the derivations given in \cite{Hatta:2011gh},
but instead 
relying on the expressions of the energy density in the gauge theory, i.e. on
the Minkowski boundary given therein.

\noindent
First consider  the quantity
\bqq\label{inttq}
 \EA = \frac{\sqrt{\lambda}}{4 \pi^2}
 \int \dif t_q\,
 \delta(\mW_q)
 \left(
 \frac{A_1}{\gq^2 \Xi^2}+
 \frac{\del}{\del t_q}\,
 \frac{A_0}{\gq \Xi^2}
 \right)~,
 \eqq
with the definitions
 \bqq\label{Wq}
 \mcal{W}_q \equiv -(t - t_q)^2 + |\br - \brq|^2,
 \qquad
 \Xi \equiv
 (t-t_q) - \bvq \cdot (\br - \brq)=
 \,\frac{1}{2}\,\frac{\dif \mcal{W}_q}{\dif t_q}\,.
\eqq
In \cite{Hatta:2011gh} the integral is evaluated by the retarded condition:
 $t_r=t_r(t, \br)$ denotes the value of $t_q$ for which $\mW_q(t_q)=0$,
with
 \bqq\label{tr}
 t - t_r = |\br - \brq(t_r)| = R~.
 \eqq
Writing $\delta(\mW_q) = \delta(t_q - t_r)/2 \vert \Xi \vert $ the result 
in the large $R$-limit  taken from \cite{Hatta:2011gh} is 

 \beq
 \EA^{ret} &=&  \frac{\sqrt{\lambda}}{8 \pi^2 R^2 \vert \xi \vert} \,
\left( \frac{\gamma^4 [\bm{a}^2 - (\bm{v} \wedge \bm{a})^2](2 -\xi)}{\xi^2} \right) 
\nn
&+&
\frac{\sqrt{\lambda}}{8 \pi^2 R^2 \vert \xi \vert} \,
\frac{\partial}{\partial t_r} 
\left( \frac{\bm{n} \cdot \bm{a} + \gamma^2 (\bm{v} \cdot \bm{a}) (2 - \xi)}{\xi^2}
\right) ~.
\label{EAret}
 \eeq

\noindent 
As a conjecture let us consider
\bqq\label{conj}
 \EA^{rad} = \frac{1}{2} (\EA^{ret} - \EA^{adv})
\eqq
by performing the integral for $\EA^{rad}$ starting from (\ref{inttq})
but using the advanced condition
with
 \bqq\label{adv}
 t - t_r =  - |\br - \brq(t_r)| = - R~.
 \eqq
This amounts to the substitutions, when  on top  
$\bm{n} \rightarrow - \bm{n}$, which does not affect the force,
\bqq
\Xi  \rightarrow -R ( 1 - \bm{n} \cdot \bm{v}) ~,
\eqq
i.e.
\bqq\label{subs}
\xi  \rightarrow - \xi~,
\eqq
whereas $\frac{\partial}{\vert \xi \vert \partial t_r}$ remains unchanged.

\noindent
From (\ref{EAret}) one obtains

 \beq
 \EA^{adv} &=&  \frac{\sqrt{\lambda}}{8 \pi^2 R^2 \vert \xi \vert} \,
\left( \frac{\gamma^4 [\bm{a}^2 - (\bm{v} \wedge \bm{a})^2](2 +\xi)}{\xi^2} \right) 
\nn
&+&
\frac{\sqrt{\lambda}}{8 \pi^2 R^2 \vert \xi \vert} \,
\frac{\partial}{\partial t_r} 
\left( \frac{ - \bm{n} \cdot \bm{a} + \gamma^2 (\bm{v} \cdot \bm{a})
 (2 + \xi)}{\xi^2}
\right)~,
\label{EAadv}
 \eeq 
and
\beq
\EA^{rad}  
 &=&
  - \frac{\sqrt{\lambda}}{8 \pi^2}\,
 \frac{\gq^4 [\bm{a}^2 - (\bm{v} \wedge \bm{a})^2]}{R^2
 \xi^2} \nn
&+& \frac{\sqrt{\lambda}}{8 \pi^2 R^2 \xi}\,
 \frac{\del}{\del t_r}
 \left[\frac{\bm{n}\cdot \baq}{\xi^2} -
\frac{\gq^2 \bm{v} \cdot \baq }{\xi}\right]~.
\label{EAA}
\eeq

\noindent
In an analogous way the contribution  $\EB^{rad}$ is evaluated,
starting from
\beq
 \EB^{ret} &=& -  \frac{\sqrt{\lambda}}{8 \pi^2 R^2 \vert \xi \vert} \,
\left( \frac{\gamma^4 [\bm{a}^2 - (\bm{v} \wedge \bm{a})^2]
(-\frac{1}{\gamma^2} +2 \xi - \xi^2)}{\xi^3} \right) 
\nn
&-&
\frac{\sqrt{\lambda}}{8 \pi^2 R^2 \vert \xi \vert} \,\frac{\del}{\del t_r}
\left( \frac{\bm{n}\cdot \baq (\xi -1)}{\xi^3}
+ \frac{\gamma^2 \bm{v} \cdot \bm{a} (2-\xi)}{\xi^2}
+  \frac{1}{\vert \xi \vert}
 \frac{\del}{\del t_r} [\frac{1}{6 \gamma^2 \xi^2} + \frac{1}{\xi}] \right)~.
\label{EBret}
\eeq
After the substitution (\ref{subs}) $\EB^{adv}$ and then $\EB^{rad}$
is obtained,
\beq
 \EB^{rad} &=& -  \frac{\sqrt{\lambda}}{8 \pi^2 R^2} \,
\left( \frac{\gamma^4 [\bm{a}^2 - (\bm{v} \wedge \bm{a})^2]
(-\frac{1}{\gamma^2}  - \xi^2)}{\xi^4} \right) 
\nn
&-&
\frac{\sqrt{\lambda}}{8 \pi^2 R^2 \xi} \,\frac{\del}{\del t_r}
\left( \frac{\bm{n}\cdot \baq }{\xi^2}
- \frac{\gamma^2 \bm{v} \cdot \bm{a} }{\xi}
+  \frac{1}{\vert \xi \vert}
 \frac{\del}{\del t_r} \frac{1}{\xi} \right)~.
\label{EBrad}
\eeq

\noindent
Finally, adding $\EA^{rad}$ and $\EB^{rad}$ the angular radiation power in 
the strong  
coupling limit is obtained
\beq\label{strongP} 
P^{rad}_{strong}(\bm{n},t) &=& P_{emitt}(\bm{n},t)
+ P_{Schott}(\bm{n},t)
\nn
&=& \frac{\sqrt{\lambda}}{8 \pi^2} ~\frac{\gamma^2
 [\bm{a}^2 - (\bm{v} \wedge \bm{a})^2]}{\xi^3}
- \frac{\sqrt{\lambda}}{8 \pi^2} \frac{d}{d t} [\frac{\bm{n} \cdot \bm{a}}{\xi^3}]~.
\eeq
The total time derivative term 
$P_{Schott}(\bm{n},t)$ differs from $P_{tt}(\bm{n},t)$ in (\ref{Ptt}).

Up to the dependence on the coupling $\lambda$ the same angular radiative
 spectrum is
found in the  weak as well as in the strong coupling limit
of the ${\cal{N}} =$~4 supersymmetric Yang-Mills theory, i.e.
$\frac{\lambda}{4} \rightarrow {\sqrt{\lambda}}$.
As in electrodynamics \cite{Schwinger} 
it is suggestive that  for strong coupling as well only the emission spectrum 
$P_{emitt}(\bm{n},t) = \frac{\sqrt{\lambda}}{8 \pi^2} ~\frac{\gamma^2
 [\bm{a}^2 - (\bm{v} \wedge \bm{a})^2]}{\xi^3}$
is
 the relevant one for radiation, i.e. for  the
 irreversible energy transfer \cite{Mikhailov:2003er}.

\noindent
Furthermore the force \cite{Chernicoff:2009re,Chernicoff:2009ff} is
\bqq\label{fstrong}
f^{\mu}_{SYM, strong} = 
\frac{\sqrt{\lambda}}{2 \pi}[ a^\nu a_\nu v^\mu + {\dot{a}}^\mu]~.
\eqq
Up to the coupling dependence 
the Abraham-Lorentz forces in classical electrodynamics as well as in the
 Yang-Mills theory have the same dependence on the acceleration $a^{\mu}$
and the velocity $v^{\mu}$, when comparing (\ref{fED}), (\ref{fpert}) 
and (\ref{fstrong}).  All do satisfy the constraint (\ref{constr}). 

As in electrodynamics \cite{Rohrlich,Fulton} the forces $f^{\mu}_{SYM,weak}$
and  $f^{\mu}_{SYM,strong}$  vanish in
weak and strong coupling  ${\cal{N}} =$~4
SYM, respectively,  for uniformly accelerated motion
 \cite{Hatta:2011gh,Xiao:2008nr}, e.g. along the $x$ direction,
$x^{\mu} = (\frac{1}{g} \sinh(g \tau), \frac{1}{g} \cosh( g \tau) =
\sqrt{t^2 + \frac{1}{g^2}}, 0,0)$,
i.e. $a_{\mu} a^{\mu} = - v_{\mu} {\dot{a}}^\mu = - g^2$ ,
although radiation is emitted.

In the spirit of \cite{Landau} a phenomenological derivation of 
$P^{rad}_{strong}$ (\ref{strongP}) could be done as follows:
retain only $P_{emitt}(\bm{n},t)$ of  (\ref{Pemitt}) as derived in 
\cite{Hatta:2011gh}, calculate after the  angular integration the force
$f^{\mu}_{emitt} = \frac{\sqrt{\lambda}}{2 \pi} a_\nu a^\nu v^\mu$.
Enforce the orthogonality (\ref{constr}) to $v^\mu$,
together with $f^{\mu}_{SYM,strong} = 0$ for uniformly accelerated motion,
 by adding the
Schott-type term $\frac{\sqrt{\lambda}}{2 \pi}{\dot{a}}^\mu$.
This one is consistently obtained after integrating
$P_{Schott}(\bm{n},t)$ assumed to be of the same from in strong as well as in
weak coupling (compare  with (\ref{weakP})).

In \cite{Maeda:2007be}  the time averaged energy density of an
oscillating quark with small linear oscillations is derived, 
$v_q(t) = \epsilon \Omega \cos{\Omega T}~, ~
a_q = -\epsilon \Omega^2 \sin{\Omega t}$ and  $\epsilon << 1$. It is
 asymptotically isotropic and
 - after correcting the numerical coefficient
   by a factor 6 - given by 
\begin{equation}
\int^{+\infty}_{-\infty}~dt <T_{00} (t, \vec{x})> =
\frac{\epsilon^2 \Omega^4 \sqrt{\lambda}}{16 \pi^2 R^2}
\int^{+\infty}_{-\infty}~dt~,
\end{equation}
which is consistent with the result
by A.~Mikhailov \cite{Mikhailov:2003er},
namely for
\begin{equation}
P_{emitt} = \frac{\sqrt{\lambda}}{2 \pi} a^2_q (t)~.
\end{equation}

\section{Conclusion}

The essence of this  note is based on the structure  of the
Abraham-Lorentz force  $f^{\mu}$ (\ref{fED}), which  holds even for
strong coupling, with the properties of the orthogonality
 to the velocity and its vanishing for uniformly accelerated motion.
It is  surprising that the force - up to its strength -
is the same in relativistic electrodynamics,
 as well as in weak and strong coupling ${\cal{N}} = $ 4~ SYM,
although the underlying angular distributions $P(\bm{n},t)$
are different.

For the ${\cal{N}} = $ 4~ SYM  model in the strong coupling limit
the special case of synchrotron radiation with frequency $\omega_0$ 
is considered in \cite{Athanasiou:2010pv}.
An independent derivation of the synchrotron radiation in this model is
given in \cite{Hubeny:2010bq}.

\noindent
In this case with $\bm{v} \cdot \bm{a} =0$, $\bm{a}^2 = v^2 \omega_0^2$
the energy density (\ref{strongP}) reads
\bqq
P^{rad}_{strong}(\bm{n},t) = \frac{\sqrt{\lambda} \omega_0^2}{8 \pi^2 \xi^4}
\left[ 3 - (4 +\gamma^{-2}) \xi - 3 v^2 \sin^2 \Theta + 2 \xi^2 \right]~,
\eqq
which differs from the one using $P_{tt}$ of (\ref{Ptt}) (compare with 
eq.~(3.71) in \cite{Athanasiou:2010pv}
and with eq.~(6.5) in \cite{Hatta:2011gh}).

\noindent
 A quantity of interest considered in \cite{Athanasiou:2010pv}
is the time-averaged angular distribution of power, given by
\bqq{\label{aver}}
\frac{d P_{emitt}(\bm{n})}{d\Omega} =  \frac{\omega_0}{2 \pi}
\int_0^{2\pi/\omega_0} ~dt ~\frac{\sqrt{\lambda}}{8 \pi^2} ~
\frac{\bm{a}^2}{(1 - \bm{n} \cdot \bm{v})^3}~,
\eqq
with  
$1 - \bm{n} \cdot \bm{v} = 1 - v \sin{\Theta} \sin{(\phi - \omega_0 t)}$.

\noindent
For this periodic motion for synchrotron radiation
the contributions of the total time derivatives
$P_{Schott}$, as well as $P_{tt}$, vanish in the time-averaged distribution.
In any case, when emitted radiation is considered total time derivative
terms should not be retained. They do not represent irreversible loss of energy
in contrast to $P_{emitt}$ \cite{Schwinger,Thirring}.

Integration in (\ref{aver}) leads to
\bqq\label{syn}
\frac{d P_{emitt}(\bm{n})}{d\Omega} = \frac{\sqrt{\lambda}}{8 \pi^2}
~\bm{a}^2 \gamma^5 
\frac{1 + \frac{v^2}{2} \sin^2{\Theta}}{(\gamma^2 \cos^2{\Theta} +
  \sin^2{\Theta})^{5/2}}~,
\eqq
which agrees with the result eq. (3.72) derived in \cite{Athanasiou:2010pv}.
The total power emitted, 
\bqq
 P_{emitt} =  \frac{\sqrt{\lambda}}{2 \pi} [\gamma^2 v \omega_0]^2~,
\eqq
is the same as the result obtained  in
 \cite{Athanasiou:2010pv,Hatta:2011gh,Mikhailov:2003er}.


\section*{Acknowledgments}

Correspondence with and helpful remarks by Dionysis N.~Triantafyllopoulos
and Krishna Rajagopal are kindly acknowledged.

\end{document}